\documentclass[slac_one]{revtex4}
\usepackage{graphicx}
\usepackage{fancyhdr}
\begin{document}

\small{
\bf {The 34th International Conference on High Energy Physics, Philadelphia, PA, USA}
}

\title{	Pixel Hit Reconstruction with the CMS Detector} 

\author{G. Giurgiu, D. Fehling, P. Maksimovic, M. Swartz }
\affiliation{The Johns Hopkins University, 3400 N. Charles Street, Baltimore, MD 21218, USA}
\author{V. Chiochia}
\affiliation{University of Zurich, Physik-Institut, Winterthurerstr. 190, Zurich, CH-8057, Switzerland}

\begin{abstract}
We present a new technique for pixel hit reconstruction with the 
CMS pixel detector. The technique is based on fitting the pixel cluster 
projections to templates obtained using a detailed simulation 
called Pixelav. Pixelav successfully describes the profiles of clusters 
measured in beam tests of radiation-damaged sensors. Originally developed 
to optimally estimate the coordinates of hits after the radiation damage,
the technique has superior performance before irradiation as well, 
reducing the resolution tails of reconstructed track
parameters and significantly reducing the light quark background of 
tagged b-quarks. It is the only technique currently available 
to simulate hits from a radiation-damaged detector.
\end{abstract}

\maketitle

\thispagestyle{fancy}

\section{THE CMS SILICON PIXEL DETECTOR} 

\begin{figure}[tbh]
\vspace{9pt}
\includegraphics[width=60mm]{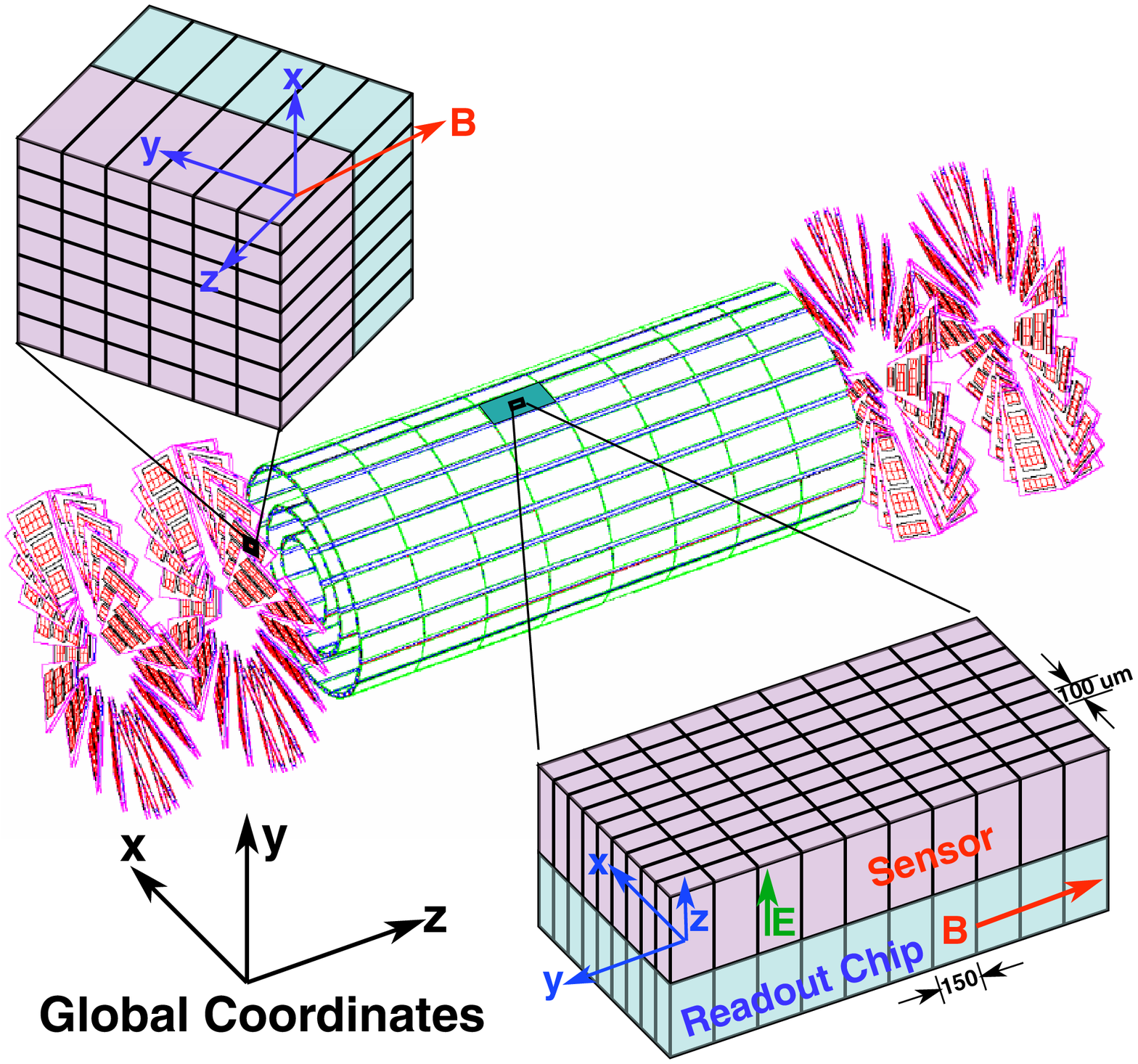}
\hspace{1cm}
\includegraphics[width=90mm]{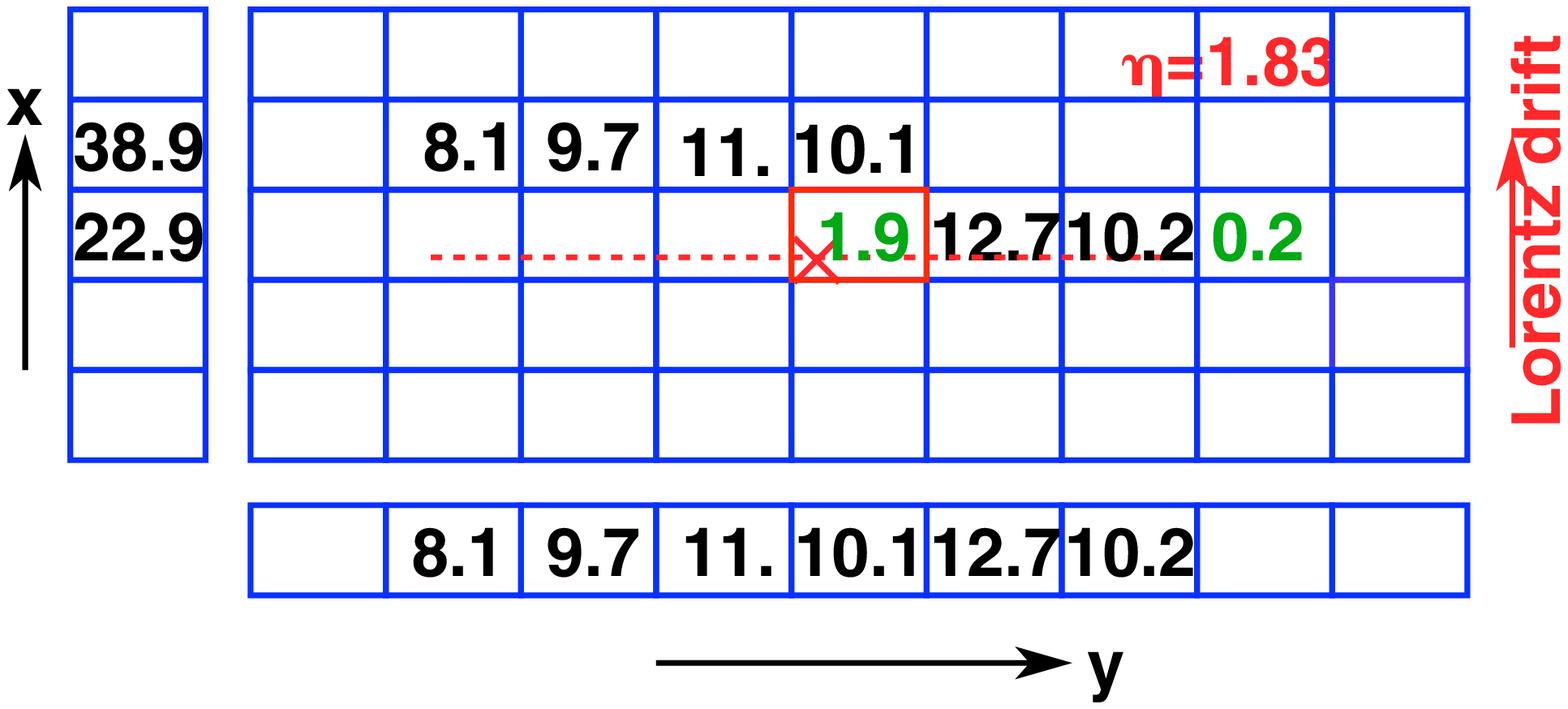}
\caption{ Left: Geometry of the CMS barrel and forward pixel detectors and the 
	corresponding coordinate systems. Right: A pixel cluster example at $\eta = 1.83$. 
	Charge deposition in each pixel is shown in thousands of electrons. 
	Numbers shown in green correspond to charge depositions below the 2000 electron 
	threshold which are not included in the cluster. The dotted red line indicates the particle trajectory projection 
	in the module plane and the red cross shows the true hit position. }
\label{fig:pixel_det}
\end{figure}

Tracking with the CMS detector is done using silicon sensors only. 
The two major components of the tracking system~\cite{Ref:track_tdr} are the silicon micro-strip 
detector and the silicon pixel detector. The micro-strip detector consists 
of barrels and end-caps and ensures coverage up to 1.2\,m radially and 
2.7\,m from the primary interaction point along the beam line. 
The pixel detector is the component closest to the beam line 
and consists of three 0.5\,m long barrel layers at radii of 4.3, 7.2 and 11\,cm 
and two end-cap disks at each end-side of the barrels located at 32.5 and 48.5\,cm 
from the interaction point. Overall, the pixel detector provides 3 hit 
coverage up to pseudo-rapidity $|\eta| < 2.5$. As shown in Figure~\ref{fig:pixel_det} 
both barrel and end-cap pixel 
detectors consist of planar detector modules mounted either on the outside radius of 
the support structure (un-flipped modules) or the inside radius of the support 
structure (flipped modules). Each silicon pixel, 150\,$\mu$m long, 
100\,$\mu$m wide and 300\,$\mu$m in depth is bump-bonded to a read-out chip.   
   
\begin{figure}[tbh]
\vspace{9pt}
\includegraphics[width=120mm]{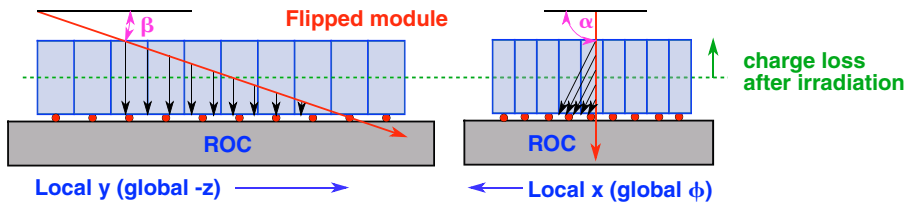}
\caption{ Charge sharing is due to either geometrical (left) or Lorentz drift (right) effects. }
\label{fig:sensors}
\end{figure}

The purpose of the tracking system is to determine the trajectories and momenta of 
charged particles produced in proton-proton collisions. In particular, the pixel 
detector is crucial for precise measurements of impact parameter and decay length of 
long lived particles like B mesons. Signatures of top quark, Higgs boson and 
super-symmetric particles include the presence of B mesons, so their identification 
based on large impact parameter or displaced decay vertex is very important.
As charged particles pass through the silicon layers, they ionize the medium 
and the charge is collected by the read-out chips. In general, the charge 
produced by a charged particle is shared by more than one pixel forming pixel 
clusters as seen in Figure~\ref{fig:pixel_det}. As shown in Figure~\ref{fig:sensors}, 
charge sharing can be due to either the particle track making a small angle with the silicon 
sensor or because of the charge drifting in the 4~Tesla magnetic field in which the 
tracking system is immersed. Before irradiation, the charge sharing is uniform in 
local X and Y directions. Even so, the reconstruction of true hit position 
is not trivial. As seen in Figure~\ref{fig:pixel_det}, due to Lorentz drift and 
threshold effects, the true hit position can be outside the observed cluster. 
Moreover, as the detector is irradiated, the defects developed 
in the silicon lattice act as charge traps so that the free charge can only partially reach  
the readout chips as schematically shown in Figure~\ref{fig:sensors}. Consequently, 
the pixel clusters become smaller, asymmetrically.       
 
\section{PIXEL HIT RECONSTRUCTION} 

We present a new technique~\cite{Ref:templates} for pixel hit reconstruction with the 
CMS pixel detector. The technique is based on fitting the pixel cluster X and Y 
projections (see Figure~\ref{fig:pixel_det}) to pre-determined cluster shapes called 
templates. A cluster template object is a map of expected charge depositions for given incidence 
angles $\alpha$ and $\beta$ defined as in Figure~\ref{fig:sensors}. 
A detailed simulation program called Pixelav~\cite{Ref:pixelav} is used to 
generate the expected cluster shapes for all possible track incidence angles $\alpha$ 
and $\beta$. The cluster template technique was 
originally developed to optimally estimate pixel hit position after radiation damage, but 
it was found that it performs better than standard CMS pixel hit reconstruction~\cite{Ref:std_reco}   
even before irradiation. This new technique requires knowledge of the track direction, so it 
is suitable for use in the second pass of pixel hit reconstruction, when track incidence angles 
on detector modules are known. In the first pass of the reconstruction the standard 
technique~\cite{Ref:std_reco} is used. 

\begin{figure}[tbh]
\vspace{9pt}
\includegraphics[width=100mm]{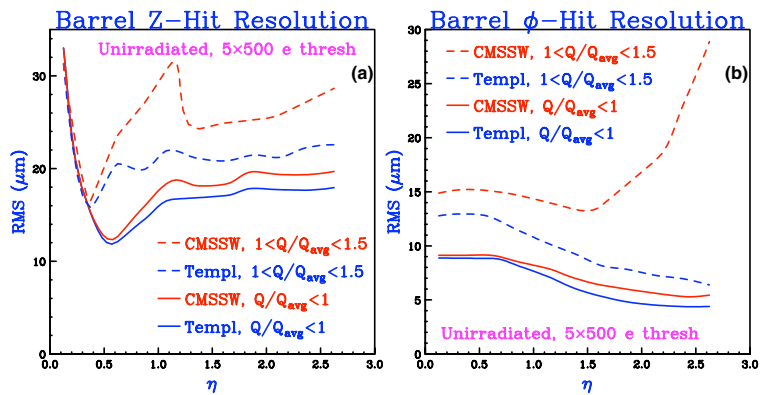}
\caption{The rms y-residuals (a) and x-residuals (b) of reconstructed barrel clusters for the 
	template (blue) and standard (red) algorithms plotted versus pseudo-rapidity for the cluster 
	charge $1.5>Q_{clus}/Q_{avg}>1$ (dashed lines) and $1>Q_{clus}/Q_{avg}$ (solid lines).  
	The sample generated by Pixelav models an un-irradiated detector operated at 150V bias.}
\label{fig:performance_no_irrad}
\end{figure}

\begin{figure}[tbh]
\vspace{9pt}
\includegraphics[width=100mm]{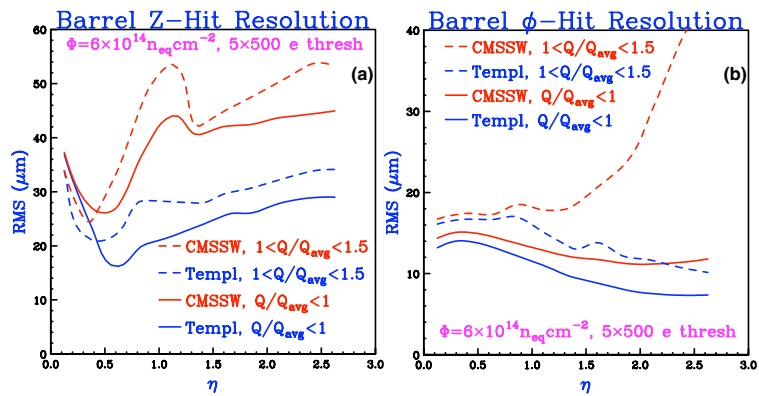}
\caption{ The rms y-residuals (a) and x-residuals (b) of reconstructed barrel clusters for the 
	template (blue) and standard (red) algorithms versus pseudo-rapidity for the clusters 
	with $1.5>Q_{clus}/Q_{avg}>1$ (dashed lines) and $1>Q_{clus}/Q_{avg}$ (solid lines). The  
	sample is generated by Pixelav with significant radiation-damage 
	($\Phi=6\times10^{14}$~n$_{\rm eq}/{\rm cm}^{2}$) operated at 300V bias.}
\label{fig:performance_irrad}
\end{figure}

Given a track with angles $\alpha$ and $\beta$ 
incident to the pixel module, the pre-determined cluster template is compared to the 
actual cluster produced by the track in question. A $\chi^2$ minimization is performed with hit 
X and Y positions as floating variables. The hit position is given by the X and Y coordinates which 
minimize the $\chi^2$ comparison.      
The template reconstruction algorithm improves the pixel hit resolution for un-irradiated sensors
as shown in Figure~\ref{fig:performance_no_irrad}.
Figure~\ref{fig:performance_irrad} shows the impressive improvement in resolution after irradiation.
Before irradiation, the improvement of the new reconstruction technique is obvious mainly 
for clusters with total charge larger than the average charge. Such clusters are affected by delta-rays which 
are accounted for in detail by the Pixelav simulation~\cite{Ref:pixelav} resulting in superior 
performance of the template technique. For clusters with charge lower than the average, the template 
and standard reconstruction techniques have similar performance. For tracks perpendicular to the 
sensor modules with $\eta \sim 0$, the Y cluster projections consist of single pixels resulting 
in poor Y resolution due to lack of charge sharing. We observe the best Y resolution around 
$\eta = 0.5$ where the Y cluster projections consist of two pixels and a slight worsening of the 
resolution with increasing $\eta$. The X resolution is significantly improved by the 
template technique for clusters with charge larger than the average and at large eta.    
After irradiation, the old reconstruction algorithm exhibits  
large biases and resolution degradation due to changes in cluster shape from charge trapping. 
The Pixelav simulation accounts for 
changes in cluster shapes due to irradiation and provides cluster templates synchronized to the 
detector irradiation state. As a result, the template reconstruction technique provides much better 
pixel hit resolution after irradiation as seen in Figure~\ref{fig:performance_irrad}.       

The significant improvement of the pixel hit resolution when using the template reconstruction 
technique leads to improvements of the track parameters, mainly the impact parameter which is 
crucial for b-tagging. We study the effect of the template reconstruction on b-tagging and find 
that the fake b-tags rates are reduced by a factor of 2-3 when the impact parameter is used 
to discriminate between true and fake b-jets. There are other applications of the pixel 
template reconstruction. The minimum $\chi^2$ value obtained from the $\chi^2$ minimization 
when templates are compared to observed clusters can be used to assess the quality of pixel hits. 
If the track direction is incompatible with the cluster shape, the $\chi^2$ probability of the 
pixel hit is low so the pixel hit can be dropped by the track pattern recognition algorithms resulting 
in significant improvement of the pattern recognition speed. Another application of the 
template technique is to re-weight the pixel clusters generated by the standard CMS simulation 
so that they agree with the observed clusters after irradiation. This way, the cluster 
simulation can be kept synchronized with the aging detector.  

We presented a new technique for pixel hit reconstruction with the CMS pixel detector. 
The method is based on simulating cluster shapes for all track incidence angles and 
different irradiation levels of the detector. The new technique improves significantly the 
pixel hit resolution, the track parameters and b-tagging fake rates and it has 
useful applications in tracking pattern recognition algorithms and simulation 
of the pixel clusters as the detector ages due to irradiation.

\end{document}